\documentclass[pra,superscriptaddress,a4paper,showpacs ]{revtex4}
\usepackage{amsmath,amsfonts,bbm, graphicx,color}

\def\>{\rangle}
\def\<{\langle}

\def\F{ \dot{{\cal F}}(E) }

\def\G{ {\cal G}^+_{\mbox{ \tiny AdS}_D } }

\def\non{ \nonumber\\}

\begin{document}

\title{On the response of a particle detector in Anti-de Sitter spacetime.} %

\author{David Jennings}%
\affiliation{Institute for Mathematical Sciences, Imperial College London, London SW7 2BW,
United Kingdom}%
\affiliation{QOLS, The Blackett Laboratory, Imperial College London, Prince Consort Road, SW7 2BW, United Kingdom}

\date{\today}

\begin{abstract}
We consider the vacuum response of a particle detector in Anti-de Sitter spacetime, and in particular analyze how spacetime features such as curvature and dimensionality affect the response spectrum of an accelerated detector. We calculate useful limits on Wightman functions, analyze the dynamics of the detector in terms of vacuum fluctuations and radiation reactions, and discuss the thermalization process for the detector. We also discuss a generalization of the GEMS approach and obtain the Gibbons-Hawking temperature of de Sitter spacetime as an embedded Unruh temperature in a curved Anti-de Sitter spacetime.
\end{abstract}
\pacs{04.62.+v, 11.10.-z, 03.75.+k }

\maketitle

\section{Introduction}

Aside from being a maximally symmetric spacetime of constant negative curvature \cite{hawking-ellis, weinberg}, Anti-de Sitter spacetime today features prominently in string theory and brane models of cosmology \cite{horava-witten, randall1}. In recent times, its importance has been further increased through the development of holography and the AdS/CFT correspondence \cite{susskind, thooft, maldacena}.

Quantum field theories on flat and curved spacetimes possess various subtleties, with one of the most celebrated results being the discovery \cite{ fulling, davies, unruh, sewell} that the Minkowski vacuum state for an accelerated observer is a thermal state, with a temperature proportional to their proper acceleration. The feature is generically called the Unruh effect, and may be explained in terms of Fock-space states where large entanglement exists across the causal horizon of the accelerated observer, resulting in them having access to a reduced thermal state.

A particularly concrete approach that uncovers the thermality of the state is for the accelerated observer to couple a discrete quantum system to the field. In this case it can be shown \cite{birrell_davies} that, to lowest order in the coupling constant $c$, the probability for transitions from the ground state $|E_0 \>$ of the particle detector, coupled to the field in its vacuum state $|0 \>$, is given by $c^2 \sum_E |\<E|M|E_0 \>|^2  {\cal F}(E)$, where $M$ is the detector observable that is coupled to the field, and we define the detector response function ${\cal F }(E)$ and the Wightman function ${\cal G}^+(x,x')$ as 
\begin{eqnarray}
{\cal F}(E) &=& \int dt dt'{\cal G}^+(x(t),x(t'))e^{-i(E-E_0)(t-t')}\nonumber\\
{\cal G}^+(x,x') &=& \<0|\phi (x) \phi (x') |0\>.
\end{eqnarray}
Consequently, if the detector response function does not vanish then excitations will occur in the detector, which correspond to the detection of particles. 

An important family of trajectories are stationary trajectories, for which the Wightman function becomes a function of $\Delta \tau =\tau - \tau '$, and where $\tau $ is the proper time along the path. However, in this case the response function diverges and instead we must consider the response function per unit time
\begin{eqnarray}
 \F &=& \int^\infty_{-\infty} d \Delta \tau e^{-i(E-E_0)\Delta \tau } {\cal G}^+(\Delta \tau ).
\end{eqnarray}  

If an accelerated observer was to probe the vacuum state by coupling a quantum system with discrete internal degrees of freedom, they would detect `clicks' as the detector is excited from its ground state. However, it has been emphasized  \cite{letaw, padmanabhan, detectorclick} that a distinction must be made between the Fock-space notion of particle content and the the particle content inferred through the clicking of a detector. While the two approaches agree in the case of a constantly accelerated observer in $4$-dimensional Minkowski spacetime, in general they differ. Indeed, states exist \cite{letaw, padmanabhan, detectorclick} in which the Fock-space analysis would conclude that the state has no particles, yet the detector clicks, and conversely states exist for which the detector never clicks, yet the Fock-space analysis concludes that there are particles in the state.

Motivated by these facts, we shall take a pragmatic approach and argue that to determine the particle content of a quantum state, ultimately we must probe the state with a particle detector that possesses internal quantum degrees of freedom, but follows a classical trajectory through spacetime. 

We can then ask: what features of the spacetime is the detector actually sensitive to? 

It is well known that for curved spacetimes, curvature-dependent responses occur. For example, a comoving detector in de Sitter has a thermal response at a temperature $T=\sqrt{ \Lambda/3}/2 \pi$, which is the Gibbons-Hawking temperature of a spacetime with constant cosmological constant $\Lambda >0$ \cite{gibbons_hawking, levin}. However, even in Minkowski spacetime subtleties arise.

The thermality of the Minkowski vacuum for accelerated observers was first calculated for the free field situation \cite{fulling, davies, unruh} and then later rigorously established for general interacting fields \cite{sewell}. The general result, due to Sewell,  shows that the KMS condition \cite{kubo, martinschwinger} holds for the Rindler wedge. However, while the KMS condition implies detailed balance and thermality, it must be emphasized that in itself this does not determine the particular shape of the observed spectrum. Following Sewell's general result, Tagaki \cite{takagi} showed explicitly for a free field in $D$-dimensional Minkowski spacetime that the response of a detector only coincides with a thermal bath response function for $D=2$ and $D=4$ \cite{takagi}, while for all other dimensions the response differs. In addition, a distinctive `inversion of statistics' occurs in odd dimensions, where a bosonic/fermionic field can display a fermionic/bosonic signature in the response function of a detector \cite{takagi, ooguri}. 

Quantum field theory on Anti-de Sitter spacetime \cite{fronsdal, fronsdal2, isham} is more challenging than on Minkowski spacetime. Indeed, it was shown \cite{levin, jacobson} that thermality arises in the free field case only for accelerations above mass scale of the Anti-de Sitter curvature. This result was later generalized in a more rigorous setting \cite{passive_vacuum1, passive_vacuum2, bros}, where in particular \cite{bros} establishes that for a general interacting field the KMS condition, and hence thermality, holds for accelerations above the mass scale. However, as with the Minkowski situation, the fact that the KMS condition holds for Anti-de Sitter does not specify the shape of the response spectrum for a particle detector in the spacetime.

The first aim of this paper is to analyze the shape of the spectrum that a particle detector in Anti-de Sitter spacetime will respond to, to see if inversion of statistics occurs and to see how the dimensionality of the spacetime and its curvature is encoded in the spectrum. 
Given this modified spectrum, our second aim will be to analyze the thermalization process in more detail and to see how vacuum fluctuations and radiation reactions each separately contribute to the process. In doing so we can explicitly show that thermalization and detailed balance are respected, despite the unusual spectrum.

Finally in the context of Anti-de Sitter spacetime we shall discuss a generalization to the GEMS approach \cite{gems1,gems2,gems3} which allows one to analyze thermal properties of a curved spacetime in terms of an Unruh response in an embedding in a higher dimensional flat spacetime.

In what follows we shall extend previously unpublished work \cite{thesis} to consider $D$-dimensional Anti-de Sitter spacetime and calculate how the response of a monopole detector is sensitive to the particular stationary classical trajectory, the dimension $D$ and also the curvature of the spacetime. The rest of the paper is organized as follows. In next section we give an overview of scalar fields in $D$-dimensional Anti-de Sitter spacetime and derive some results concerning Wightman functions in the spacetime. In section \ref{response} we extend some existing work \cite{levin}, calculate analytical expressions for the response function of a detector and show how it depends on the particular acceleration, on the spacetime dimension, and on the curvature. We find that `inversion of statistics' occurs in a way similar to Minkowski spacetime. Then in section \ref{vacrad} we discuss how the excitations of the detector can be analyzed in terms of radiation reactions and vacuum fluctuations and describe the  thermalization process of the detector. In section \ref{GHtemp} we discuss how by constraining to de Sitter sub-manifolds that the Gibbons-Hawking temperature naturally emerges, which is a generalization of the Global Embedding Minkowski Spacetime (GEMS) approach \cite{gems1, gems2, gems3} to curved spacetimes. We then conclude with section \ref{conclude}.

\section{Scalar fields in D-dimensional Anti-de Sitter Spacetime}

Quantum field theory in Anti-de Sitter spacetime possesses certain delicate features, such as its lack of global hyperbolicity \cite{hawking-ellis, fronsdal, fronsdal2, isham}. Despite such difficulties it has been argued that, under reasonable boundary conditions, thermality of its vacuum state still arises, but only for accelerations above a critical acceleration of $\sqrt{|\Lambda|/3}$. In particular, this result has been argued in \cite{levin}, where they consider the 4-dimensional thermal responses in both de Sitter and Anti-de Sitter spacetime, and also shown to hold in a more general field theory framework  \cite{passive_vacuum1, passive_vacuum2, bros}.

It is necessary to derive some results that are needed in order to deal with a detector coupled to any Anti-de Sitter fields. Since, among other things, we want to investigate the dependence on spacetime dimension, we consider a general $D$-dimensional Anti-de Sitter spacetime with constant Ricci scalar $R=-D(D-1)k^2$. Here $k$ is the curvature of the spacetime and related to the negative cosmological constant through $| \Lambda| = k^2(D-1)(D-3)/2$.  The metric for this spacetime may be expressed in Poincar\'{e} coordinates as
\begin{eqnarray}
 ds^2&=&e^{-\frac{2y}{k}}( dt^2 - dx_1^2 - dx_2^2 \cdots -dx_{D-2}^2) -dy^2,
\end{eqnarray}
which is made conformally flat by the change of variables $dz=e^{\frac{y}{k}}dy$ to produce
\begin{eqnarray}
ds^2&=&\frac{1}{k^2 z^2}( dt^2 - dx_1^2 - dx_2^2 \cdots -dx_{D-2}^2 -dz^2).
\end{eqnarray}
The equation of motion of a scalar field of mass $m$, coupled to gravity in this spacetime, is given by
\begin{eqnarray}
(\nabla^\mu \nabla_\mu +m^2 +\zeta R) \phi(x) &=&0, 
\end{eqnarray}
where $\zeta$ is its coupling to gravity.

Making use of the planar symmetry, it can be seen that modes exist of the form
\begin{eqnarray}
 \phi(x)&=& \phi_{\textbf{p}}(x^i)\beta_n(y)\nonumber \\
 \phi_{\textbf{p}}(x^i)&=& \frac{e^{-i\eta_{ij}p^ix^j}}{\sqrt{2\omega(\textbf{p}) (2\pi)^{D-2}}},
\end{eqnarray}
where $\eta_{ij}$ is the metric for an $(D-1)$-dimensional Minkowski spacetime and 
\begin{eqnarray}
 p^i&=&(\omega(\textbf{p}), \textbf{p}) \nonumber\\
\omega(\textbf{p}) &=& \sqrt{|\textbf{p}|^2 + m_n^2}.
\end{eqnarray}
The constants $m_n$ are obtained from a separation of variables in the equation of motion and are determined once boundary conditions are specified, while the functions $\beta_n(y)$, which describe the field transverse to the Minkowski planes, are given by
\begin{eqnarray}
 \beta_n(y)&=&c_n e^{(D-1)\frac{ky}{2}}(J_\nu(m_nz) + b_\nu Y_\nu(m_nz))
\end{eqnarray}
where $J_\nu(m_nz)$ and  $ Y_\nu(m_nz)$ are Bessel and Neumann functions respectively.

The parameter $\nu $ is of importance and behaves like an effective scale for the field. It is given by
\begin{eqnarray}
 \nu &=& \sqrt{\frac{(D-1)^2}{4} -D(D-1)\zeta +\frac{m^2}{k^2}}.
\end{eqnarray}

\subsection{Wightman functions for a Scalar field in Anti-de Sitter}\label{wightm}
The coupling of the scalar field to a detector moving in this spacetime is governed by a two-point function of the field. More specifically, when the field is in a vacuum state the Wightman function provides the necessary information. The Wightman function for a scalar field in a vacuum state $|0\>$, is defined as ${\cal G}^+(x,x')=\<0|\phi(x)\phi(x')|0\>$.
For the case of Anti-de Sitter spacetime, the Wightman function for suitable boundary conditions can be calculated \cite{fronsdal, fronsdal2, ads_wightman} to be of the form 
\begin{eqnarray}
 {\cal G}^+(x,x')&=&k^{D-2} (zz')^{\frac{D-1}{2}} \int_0^\infty dm \left [m{\cal G}^+_{{\mathbb M}_{D-1}}(x,x';m)J_\nu(mz)J_\nu(mz')\right ], 
\end{eqnarray}
where ${\cal G}^+_{{\mathbb M}_{D-1}}(x,x';m)$ is the Wightman function for a scalar field, of mass $m$, in an $(D-1)$-dimensional Minkowski spacetime. The explicit form of this lower dimensional Wightman function is
\begin{eqnarray}
 {\cal G}^+_{{\mathbb M}_{D-1}}(x,x';m)&=& \frac{m^{\alpha-\frac{1}{2}}K_{\alpha -\frac{1}{2}} (m\sqrt{(\textbf{x}-\textbf{x}')^2 - (t-t'-i \epsilon )^2})}{(2\pi)^{\alpha +\frac{1}{2}}((\textbf{x}-\textbf{x}')^2 - (t-t'-i \epsilon )^2)^{\frac{\alpha}{2} -\frac{1}{4}}}
\end{eqnarray}
where $K_\nu(x)$ is a modified Bessel function, $\epsilon $ is the regularization, and we have introduced for convenience $\alpha =\frac{D}{2}-1$.

We then find that the Wightman function for a massive scalar field in $D$-dimensional Anti-de Sitter spacetime is given by
\begin{eqnarray}\label{wightman}
  {\cal G}^+_{AdS_{D}}(x,x')&=& \frac{k^{2 \alpha }e^{-\alpha \pi i}}{(2\pi)^{\alpha +1}}\frac{1}{(v^2-1)^{\frac{\alpha} {2}}}Q^\alpha_{\nu -\frac{1}{2}} (v)
\end{eqnarray}
where
\begin{eqnarray}\label{v}
 v&=&\frac{z^2 + z'^2 + (\textbf{x} -\textbf{x}')^2 -(t-t'-i \epsilon )^2}{2zz'}, 
\end{eqnarray}
and $Q^\alpha_\beta (x)$ is the associated Legendre function of the second kind.

\subsection{Limits on mass and curvature for the Anti-de Sitter Wightman Function}\label{wightm2}
The form of the Wightman function for a massive scalar field is relatively awkward to use, and so in this section we establish some analytic limits for $\G (x,x')$. In particular, we shall see that the corresponding massless conformally coupled case takes on a much more flexible form. We also check our results by computing the two different limiting routes by which we may pass to a massless field in $D$-dimensional Minkowski spacetime.
\subsubsection{Massless scalar field in Minkowski spacetime}
To calculate the Wightman function for a massless scalar field in the Minkowski spacetime we may first reduce the curvature of the spacetime to zero and then allow the mass of the field to vanish.

It can be shown (see e.g. \cite{ads_wightman}) that (\ref{wightman}) tends to ${\cal G}^+_{{\mathbb M}_D}$ in the limit $k\rightarrow 0$. Namely,
\begin{eqnarray}
\lim_{k\rightarrow 0} {\cal G}^+_{AdS_D}(x,x') &=&{\cal G}^+_{{\mathbb M}_D}(x,x')= \frac{m^\alpha}{(2\pi)^{\alpha +1}w^\alpha }K_\alpha (mw), 
\end{eqnarray}
where $w^2 = (y-y')^2 +(\textbf{x} +\textbf{x}')^2 -(t-t'-i \epsilon )^2$ and $K_\alpha(x)$ is the modified Bessel function.

To compute the massless limit of this we observe that for small $x$
\begin{eqnarray}
 K_\alpha (x)&\sim& \frac{1}{2} \Gamma(\alpha )\left ( \frac{2}{x}\right )^\alpha  
\end{eqnarray}
from which we recover \cite{birrell_davies} the standard Minkowski result
\begin{eqnarray}
\lim_{m \rightarrow 0} \lim_{k \rightarrow 0}\G (x,x') &=& \frac{\Gamma (\alpha )}{4 \pi^{\alpha +1} w^{2 \alpha }}.
\end{eqnarray}

\subsubsection{Massless scalar field in Anti-de Sitter spacetime}

To compute the Wightman function for a massless scalar field in Anti-de Sitter spacetime we need a well behaved expansion for $Q^\alpha _\beta (x)$ in the limit of vanishing mass. One such expansion \cite{abramowitz} is
\begin{eqnarray}
 Q^\alpha _\beta (x)&=&\frac{e^{\alpha \pi i}}{2^{\beta +1}}\frac{\Gamma( \alpha + \beta + 1)}{\Gamma(\beta +1)}(x^2 -1)^{\frac{\alpha}{2}} \int^1_{-1}\frac{(1-t^2)^\beta}{ (x -t)^{\alpha + \beta + 1}}dt.
\end{eqnarray}
With this we find that
\begin{eqnarray}
 \lim_{m \rightarrow 0} {\cal G}^+_{AdS_D}(x,x')&=& \frac{k^{2 \alpha }}{(2 \pi )^{\alpha +1}2^{\nu_0 +\frac{1}{2}}}\frac{\Gamma( \alpha + \nu_0 +\frac{1}{2})}{\Gamma (\nu_0 +\frac{1}{2})} \int^1_{-1}\frac{(1-t^2)^{\nu_0 -\frac{1}{2}}}{(v-t)^{\alpha + \nu_0 +\frac{1}{2}}}dt 
\end{eqnarray}
where $\nu_0 = \sqrt{\frac{(D-1)^2}{4} -D(D-1)\zeta}$. 
We now make the further assumption that the massless scalar field is conformally coupled to gravity, and so $\nu_0 =\frac{1}{2}$. We conclude that
\begin{eqnarray}\label{G}
  \lim_{m \rightarrow 0} {\cal G}^+_{AdS_D}(x,x')&=&{\cal G}^+_{AdS_D,m=0}(x,x')\nonumber \\ 
&=& \frac{k^{2 \alpha } \Gamma (\alpha)}{2(2\pi)^{\alpha +1}} \left ( \frac{1}{(v-1)^ \alpha } - \frac{1}{(v+1)^ \alpha }  \right ).
\end{eqnarray}

In particular, this result behaves correctly as we shrink the curvature to zero. For small $k$, $v \sim 1 +\frac{k^2w^2}{2}$ and consequently
\begin{eqnarray}
{\cal G}^+_{AdS_D,m=0}(x,x') &\sim&  \frac{k^{2 \alpha } \Gamma (\alpha)}{2(2\pi)^{\alpha +1}} \left ( \frac{2^\alpha }{k^{2 \alpha}w^{2 \alpha } } - \frac{1}{(2 + \frac{k^2w^2}{2})^ \alpha }\right )
\end{eqnarray}
Taking the limit as the curvature tends to zero gives
\begin{eqnarray}
 \lim_{k\rightarrow 0}  \lim_{m \rightarrow 0} \G (x,x')&=&\frac{\Gamma (\alpha )}{4 \pi^{\alpha +1} w^{2 \alpha }},
\end{eqnarray}
in agreement with our previous result on the massless limit of the Minkowski Wightman function.

\section{Unruh response for a detector in Anti-de Sitter Spacetime}\label{response}

In this section we consider a detector, coupled to a massless conformally coupled scalar field, that is moving along certain stationary trajectories in the Anti-de Sitter spacetime. The most basic stationary trajectories are the timelike geodesic, and the constant acceleration trajectories. We shall also find that excitations only occur above a certain critical acceleration. We shall find that `inversion of statistics' occurs where it exhibits a fermionic signature in odd dimensions and a bosonic signature in even dimensions. The relevant Wightman function for such a field is given by (\ref{G}) and for simplicity we take the ground state energy $E_0$ to be zero. 

\subsection{Trajectories of constant acceleration}\label{wight}

\subsubsection{Geodesics}
We work with the conformally flat metric
\begin{eqnarray}
ds^2&=&\frac{1}{k^2z^2}( dt^2 - dx_1^2 - dx_2^2- \cdots -dx_{D-2}^2 -dz^2),
\end{eqnarray}
and for simplicity restrict motion to the $z-t$ plane. The timelike geodesics can be written as
\begin{eqnarray}
 \gamma k t &=& \tan (k \tau + A )\nonumber\\
 \gamma k z &=& \sec (k \tau+ A),
\end{eqnarray}
where $\gamma$ and $A$ are constants and $\tau$ is the proper time. 

We wish to consider ${\cal G}^+(x(\tau),x(\tau '))$ at two points along the geodesic. We find that $v(\tau,\tau ')=\cos(k \Delta \tau - i \epsilon)$, and so substitution into (\ref{G}) then gives that
\begin{eqnarray}
 {\cal G}^+(\Delta \tau)&=& \frac{k^{D-2 } \Gamma(\frac{D}{2}-1)}{2(2\pi)^{\frac{D}{2}}} \left [ \frac{1}{(\cos (k \Delta \tau -i \epsilon )  -1)^{\frac{D}{2}-1} }- \frac{1}{(\cos (k \Delta \tau -i \epsilon )  +1)^{\frac{D}{2} -1} } \right ].
\end{eqnarray}

All the poles of this function lie in the upper half plane, at the points $\frac{n \pi}{k} + i \epsilon $, $n \in {\mathbb Z}$. Furthermore, since $E>0$ the integral
\begin{eqnarray}
\F = \int^\infty_{-\infty} e^{-iE\Delta \tau} {\cal G}^+(\Delta \tau) d \Delta \tau
\end{eqnarray}
may be evaluated by closing a contour in the lower half plane. Since there are no poles in this region we deduce that the integral vanishes. No excitations are observed, as is expected for a geodesic path.

We now turn to non-intertial trajectories. The most straightforward of these trajectories are those where the acceleration remains constant throughout. By considering planar intersections with the spacetime in a flat embedding space it is possible to obtain all constant acceleration curves \cite{bros}. These curves are then conic sections of the AdS spacetime. There are three classes of constant acceleration timelike curves: elliptic, parabolic and hyperbolic. Elliptic trajectories are those curves for which the acceleration, $a$, satisfies $a^2 < k^2$, parabolic curves are those that satisfy $a^2=k^2$, while hyperbolic trajectories satisfy $a^2 >k^2$. The acceleration is then said to be subcritical, critical or supercritical respectively.

\subsubsection{Subcritical accelerations}
Trajectories with accelerations below $k$ in the Poincar\'{e} coordinate system can be described by
\begin{eqnarray}
 t( \tau )&=& z_0 k \sinh \gamma (\tau ) \nonumber\\ 
 z( \tau )&=& z_0 k \cosh \gamma (\tau )+z_0a  ,
\end{eqnarray}
with
\begin{eqnarray}
 \cosh \gamma ( \tau )&=& \frac{a \cos( \sqrt{k^2 -a^2} \tau ) - k}{a- k \cos ( \sqrt{k^2 -a^2} \tau )}.
\end{eqnarray}
Evaluating at two points along the trajectory gives us 
\begin{eqnarray}
 v( \tau , \tau ')&=&\frac{k^2}{k^2 -a^2} \cos( \sqrt{k^2 -a^2} \Delta \tau - i \epsilon ) 
\end{eqnarray}  
which in turn gives 
\begin{eqnarray}
 {\cal G}^+(\Delta \tau)&=& \frac{k^{D-2 } \Gamma(\frac{D}{2}-1)}{2(2\pi)^{\frac{D}{2}}} \left [ \frac{1}{(\frac{k^2}{k^2 -a^2} \cos( \sqrt{k^2 -a^2} \tau  - i \epsilon ) -1)^{\frac{D}{2}-1}} -\frac{1}{(\frac{k^2}{k^2 -a^2} \cos( \sqrt{k^2 -a^2} \tau  - i \epsilon ) +1)^{\frac{D}{2}-1}} \right ].
\end{eqnarray}
Once again, no poles exist in the lower half plane, $\F=0$ and so no excitations from the ground state are detected for subcritical accelerations.
\subsubsection{Critical accelerations}
Trajectories with critical accelerations may be described as
\begin{eqnarray}
 z&=&z_0\nonumber \\
t &=& kz_0\tau  .
\end{eqnarray}
For two points along such a trajectory with have $v(\tau,\tau')=1- \frac{k^2}{2} ( \Delta \tau -i \epsilon )^2$ and so
\begin{eqnarray}
 {\cal G}^+(\Delta \tau)&=& \frac{k^{D-2 } \Gamma(\frac{D}{2}-1)}{2(2\pi)^{\frac{D}{2}}}\left [ \frac{2^{D/2-1}}{k^{D-2} (\Delta \tau - i \epsilon )^{D-2}} - \frac{1}{(2 - \frac{k^2}{2}(\Delta \tau -i \epsilon )^2)^{\frac{D}{2}-1}}  \right ].
\end{eqnarray}
It is clear that once again all singularities lie in the upper half plane, and so despite being accelerated, no excitations are detected. We also note that the first term is identical to the result obtained for a \textit{geodesic} trajectory in $D$-dimensional \textit{Minkowski} spacetime. 

\subsubsection{Supercritical accelerations}
The supercritical trajectories are given by
\begin{eqnarray}
 t&=& \frac{az_0}{\omega }e^{\omega \tau}\nonumber \\
 z&=& z_0e^{\omega \tau},
\end{eqnarray}
with $\tau$ once again being the proper time along the trajectory and $\omega =\sqrt{a^2 -k^2}$.

By considering two points along a given trajectory we obtain
\begin{eqnarray}
 v(\tau, \tau ')&=&\frac{a^2}{\omega ^2} -\frac{ k^2}{\omega ^2} \cosh(\omega \Delta \tau - i \epsilon ).
\end{eqnarray}
Substituting this into (\ref{G}) gives us that the Wightman function along the trajectory is now
\begin{eqnarray}
{\cal G}^+(\Delta \tau)&=& \frac{\omega ^{D-2} \Gamma (\frac{D}{2}-1)}{(4\pi)^{\frac{D}{2}}} \left [ \frac{1}{i^{D-2}\sinh ^{D-2}(\frac{\omega \Delta \tau}{2} -i \epsilon)} - \frac{1}{(\sinh(A + (\frac{\omega \Delta \tau}{2} -i \epsilon) ))^{\frac{D}{2}-1}  (\sinh(A - (\frac{\omega \Delta \tau}{2} -i \epsilon) ))^{\frac{D}{2}-1}}  \right ] \nonumber
\end{eqnarray}
with $\sinh A = \frac{\omega}{ k}$.

The second term does not feature in the limit to Minkowski spacetime, and is purely from the non-zero curvature that the spacetime possesses.

We calculate the detector response function per unit time for this path by first dividing the Wightman function into two terms, ${\cal G}^+(\Delta \tau) = {\cal G}^+_1(\Delta \tau)- {\cal G}^+_2(\Delta \tau)$, and calculate the contribution from each term separately. 
\begin{eqnarray}
{\cal G}^+_1(\Delta \tau)&=& \frac{\omega ^{D-2} \Gamma (\frac{D}{2}-1)}{i^{D-2}(4\pi)^{\frac{D}{2}}} \left [ \frac{1}{\sinh ^{D-2 }(\frac{\omega \Delta \tau}{2} -i \epsilon)} \right ], \nonumber\\
{\cal G}^+_2(\Delta \tau)&=& \frac{\omega ^{D-2} \Gamma (\frac{D}{2})}{(4\pi)^{\frac{D}{2}}} \left [\frac{1}{(\sinh(A + (\frac{\omega \Delta \tau}{2} -i \epsilon) ))^{\frac{D}{2}-1}(\sinh(A - (\frac{\omega \Delta \tau}{2} -i \epsilon) ))^{\frac{D}{2}-1} }  \right ].
\end{eqnarray}
 We first evaluate the contribution due to ${\cal G}^+_1$.  The integrand of
\begin{eqnarray}
 \int^\infty_{-\infty} e^{-iE \Delta \tau} {\cal G}^+_1(\Delta \tau) d \Delta \tau = \frac{\omega ^{D-2} \Gamma (\frac{D}{2}-1)}{i^{D-2}(4\pi)^{\frac{D}{2}}}\int^\infty_{-\infty} \frac{e^{-iE \Delta \tau}}{\sinh ^{D-2} (\frac{\omega \Delta \tau}{2} -i \epsilon )} d \Delta \tau 
\end{eqnarray}
has poles of order $(D-2)$ all along the imaginary axis at the points $\Delta \tau = i( \epsilon -\frac{2m\pi}{ \omega} )$, with $m \in {\mathbb Z}$. The residues may be calculated by standard methods and we find that the contribution to the response function is given by
\begin{eqnarray}
 \int^\infty_{-\infty}e^{-iE \Delta  \tau}{\cal G}^+_1(\Delta \tau) d \Delta \tau&=& \frac{\Gamma(\frac{D}{2}-1) \omega ^{D-3} f_D(\frac{E}{ \omega} )}{(4 \pi )^{\frac{D}{2}}}\frac{1}{e^{2 \pi \frac{E}{ \omega} } - (-1)^D}
\end{eqnarray}
where $f_D(\frac{E}{\omega })$ is a real polynomial of order $D-3$. Explicitly, for $D=4,5,6$ we have the expressions
\begin{eqnarray}
f_4\left (\frac{E}{\omega} \right ) &=&8\pi \frac{E}{\omega} \nonumber \\ 
f_5\left (\frac{E}{\omega} \right ) &=&2\pi\left (1+\frac{E^2}{\omega^2} \right )\nonumber \\ 
f_6 \left ( \frac{E}{\omega} \right ) &=&\frac{16\pi}{3} \left ( \frac{E}{\omega}+ \frac{E^3}{\omega^3} \right ).
\end{eqnarray}
We also note the fermionic signature in odd dimensions.

The contribution due to ${\cal G}^+_2$ can be dealt with in a similar way. In this case the function is slightly more complicated. The function
\begin{eqnarray}
{\cal G}^+_2(\Delta \tau)&=& \frac{\omega ^{D-2} \Gamma (\frac{D}{2}-1)}{(4\pi)^{\frac{D}{2}}} \left [\frac{1}{(\sinh(A + (\frac{\omega \Delta \tau}{2} -i \epsilon) ))^{\frac{D}{2}-1} (\sinh(A - (\frac{\omega \Delta \tau}{2} -i \epsilon) ))^{\frac{D}{2}-1} }  \right ] 
\end{eqnarray}
has poles along two lines in the complex plane. Specifically for $m \in {\mathbb Z}$, it is singular at 
\begin{eqnarray}
\Delta \tau_{m, \pm} = i \epsilon \pm \frac{2A}{\omega } -\frac{2m\pi i}{\omega }.
\end{eqnarray}
For even dimensional spacetimes these poles are isolated and the residue theorem may be employed. Calculation of the residue at $\Delta \tau_{m,+}$ plus the residue at $\Delta \tau_{m,-}$ gives 
\begin{eqnarray}
 \frac{\Gamma (\frac{D}{2}-1)ie^{-2m\pi \frac{E}{\omega} }}{2\pi(4\pi)^{\frac{D}{2}}} k^{D-2}( p_D(E,a,k) \sin \theta + q_D(E,a,k) \cos \theta )
\end{eqnarray}
where $\theta = \frac{2E}{\omega }\sinh^{-1} \left ( \frac{ \omega }{k} \right )$, and $p_D$ and $q_D$ are functions analogous to $f_D$. The full contribution from ${\cal G}^+_2$ then becomes 
\begin{eqnarray}
  \int^\infty_{-\infty}  e^{-iE \Delta \tau} {\cal G}^+_2(\Delta \tau)d \Delta \tau&=& \frac{k^{D-2} \Gamma(\frac{D}{2}-1)  }{(4 \pi )^{\frac{D}{2}}}\frac{p_D(E,a,k) \sin \theta + q_D(E,a,k) \cos \theta}{e^{2 \pi \frac{E}{ \omega} } - 1}. 
\end{eqnarray}

Where, for example we have 
\begin{eqnarray}
p_4(E,a,k) &=&\frac{4\pi}{a},\hspace{2.1cm} q_4(E,a,k)=0  \nonumber \\
p_6(E,a,k) &=&4\pi \left (\frac{1}{a} - \frac{k^2}{2a^3} \right ),\hspace{0.3cm} q_6(E,a,k)=-\frac{4 \pi E}{a^2}.  
\end{eqnarray}
By increasing the spacetime dimension, $p_D$ and $q_D$ become higher order polynomials in $\frac{1}{a}$.

 For odd dimensions branch points exist in ${\cal G}_2^+$ and evaluation of the second term becomes more awkward to work with, however we can still analyze thermality from the fact that ${ \cal G} _2^+(\Delta \tau + \frac{2 \pi i}{\omega} )= (-1)^D  { \cal G }_2^+(\Delta \tau )$, and also ${ \cal G} _1^+(\Delta \tau + \frac{2 \pi i}{\omega} )= (-1)^D  { \cal G }_1^+(\Delta \tau )$. Consequently for odd dimensions we then have ${ \cal G}^+(\Delta \tau + \frac{2 \pi i}{\omega} )= -{ \cal G}^+(\Delta \tau)$, which is the KMS condition of thermality for a fermionic field. This condition ensures that detailed balance holds for the detector, and furthermore results in a variation of statistical distribution $(e^{E/kT}- (-1)^D)^{-1}$  for the particle detector, in $D$-dimensions. A detailed analysis of this alternating of statistics can be found in \cite{ooguri}. We obtain the overall response by adding together the two contributions from ${ \cal G}_1^+$ and ${ \cal G}_2^+$, and obtain the response rate
\begin{eqnarray}
\F = \int^\infty_{-\infty} d \Delta \tau {\cal G}^+(\Delta \tau)&=&\frac{F(E,a,k)}{e^{2 \pi \frac{E}{ \omega} } - (-1)^D},
\end{eqnarray}
where the function $F(E,a,k)$ is given by
\begin{eqnarray}
 F(E,a,k)&=& \frac{\Gamma (\frac{D}{2}-1)}{(4 \pi )^{\frac{D}{2}}} (\omega ^{D-3}f_D(\frac{E}{\omega }) - k^{D-2}g_D(E,a,k)),
\end{eqnarray}
and both $f_D$ and $g_D$ are obtained from a contour integral around singular points in the lower half plane of complex $\Delta \tau $. Explicitly, in four dimensional Anti-de Sitter we have
\begin{eqnarray}
\F =  \left( \frac{E}{2 \pi} - \frac{k^2}{4 \pi a} \sin \left [\frac{2E}{\omega} \sinh^{-1} \left (\frac{\omega}{k} \right ) \right ] \right )  \frac{\Theta(a-k)}{e^{2 \pi E/\omega} -1} ,
\end{eqnarray}
while in six dimensional Anti-de Sitter we have
\begin{eqnarray}
\F =  \left( \frac{E^3+E\omega^2}{12 \pi^2} - \frac{k^4}{16 \pi^2 a^2} \left [ \frac{(2a^2-k^2)}{2a}\sin \left[\frac{2E}{\omega} \sinh^{-1} \left(\frac{\omega}{k} \right ) \right] -E\cos \left [\frac{2E}{\omega} \sinh^{-1} \left ( \frac{\omega}{k} \right ) \right] \right ] \right )  \frac{\Theta(a-k)}{e^{2 \pi E/\omega} -1} 
\end{eqnarray}
where $\Theta(x)$ is the Heaviside function.

We conclude that an observer with a detector, that has a constant acceleration in an Anti-de Sitter spacetime with curvature $k$ will register no Unruh response until his acceleration exceeds the curvature scale of the spacetime. Once he exceeds this threshold acceleration the response of the detector is quasi-thermal, meaning it obeys the KMS condition at a temperature
\begin{eqnarray}
 T&=&\frac{\sqrt{a^2 - k^2}}{2 \pi },
\end{eqnarray}
but the response spectrum is modified by both the dimensionality of the spacetime and by the curvature, which causes a weak oscillatory signature in all dimensions, as described by the functions $p_D(E,a,k)$ and $q_D(E,a,k)$.

We have also observed `inversion of statistics' for the supercritical accelerations, and found that for a fixed value of acceleration and curvature, increasing the dimension of the spacetime tends to enhance the response rate of the detector, through a polynomial $f_D(\frac{E}{\omega})$.

\section{Vacuum Fluctuations and Radiation Reactions}\label{vacrad}

We now analyze in more detail the detector excitations due to the non-inertial motion of the detector through the Anti-de Sitter spacetime. To do so we use a model of a detector coupled to a scalar field in the spacetime, and follow previous work done for flat spacetime \cite{accelerated_atom} and more recently for some curved spacetimes \cite{atom-deSitter, atoms-schwartz} that makes use of a well-defined separation of detector excitations into those that stem primarily from fluctuations in the vacuum, and those that are due to the disturbance of the scalar field resulting from the matter coupling. In particular, this approach is useful for us to examine the approach to equilibrium of the particle detector to ensure that it achieves genuine thermal equilibrium with the quantum field at the correct temperature, independent of the dimension or curvature of the spacetime.

We assume that the particle detector has no internal spatial variations, discrete multiple energy levels $\{ E_i \}$ and that it is coupled to a scalar field $\phi (x)$ by a term $c M(\tau) \phi(x(\tau))$, where $\tau$ is the proper time along the trajectory of the detector. We take the Hamiltonian for the whole system to be
\begin{eqnarray}
 H_{\mbox{\tiny total \normalsize}}&=& H(\tau) + \int d^3k \frac{dt}{d \tau} a^\dagger(\textbf{k},t)a(\textbf{k},t)\omega( \mathbf{k}) + cM(\tau) \phi (x(\tau)),
\end{eqnarray}
with $H(\tau)$ the Hamiltonian for the detector and $\{a(\mathbf{k},t)\}$ a mode decomposition for the scalar field.

We now make a well-defined division of each physical observable $A(\tau)=A_0(\tau) + A_s(\tau)$ into a \emph{free} part $A_0(\tau)$ and a \emph{source} part $A_s(\tau)$ with the property that $\lim_{c\rightarrow 0} A(\tau) = A_0(\tau)$, and $\<A_s (\tau)\>$ is first order in $c$ on any state.

We may simply define the source part of $A(\tau)$ through first order perturbation as
\begin{eqnarray}
A_s(\tau) &=& ic \int^\tau_0 d \tau ' \phi_0(x(\tau '))[M_0(\tau '),A_0(\tau )] 
\end{eqnarray}
for a detector observable, and for a field observable along the trajectory
\begin{eqnarray}
A_s(x(\tau)) &=& ic \int^\tau_0 d \tau 'M_0(\tau ') [\phi_0(x(\tau ')),A_0(\tau )] .
\end{eqnarray}

For a detector observable $A(\tau)$ we may decompose its evolution into contributions coming from the unperturbed free part of the field and from the perturbed source part of the field, so that $\dot{A}(\tau) =\dot{A}(\tau)_{\mbox{\tiny vac}}+ \dot{A}(\tau)_{\mbox{\tiny rad}}$. The contributions are the vacuum fluctuations and radiation reactions respectively.

A subtlety arises when we attempt to perform this decomposition. To arrive at a physically unambiguous and well-defined notion of vacuum fluctuations as separate from radiation reactions it is necessary \cite{DDC} to use a symmetric form of the operator ordering. The vacuum fluctuations and radiation reactions are then given, in a well-defined manner, by the expressions
\begin{eqnarray}
\dot{A} (\tau)_{\mbox{\tiny vac \normalsize}}&=& i \frac{c}{2} \left [ \phi_0(x(\tau))[M(\tau),A(\tau)] + [M(\tau),A(\tau)] \phi_0(x(\tau))\right ]\nonumber\\
\dot{A} (\tau)_{\mbox{\tiny rad \normalsize}}&=& i \frac{c}{2} \left [ \phi_s(x(\tau))[M(\tau),A(\tau)] + [M(\tau),A(\tau)] \phi_s(x(\tau))\right ].
\end{eqnarray}
It is then possible to expand to ${\cal O}(c^2)$, assuming a vacuum state for the field, and deduce that the contributions to the evolution of the detector's Hamiltonian that come from vacuum fluctuations and radiation reactions are given by the expressions 
\begin{eqnarray}
\<0,E_i|\dot{H}(\tau)_{\mbox{\tiny vac }}|0,E_i\>&=& 2ic^2\int^\tau_0 d\tau 'C_\phi(x(\tau) , x(\tau'))\frac{d}{d\tau}\chi_i (\tau, \tau') \nonumber\\
\<0,E_i|\dot{H}(\tau)_{\mbox{\tiny rad \normalsize}}|0,E_i\>&=& 2ic^2\int^\tau_0 d\tau '\chi_\phi (x(\tau ) , x(\tau'))\frac{d}{d\tau}C_i (\tau, \tau')
\end{eqnarray}
where we have introduced the two-point functions along the trajectory $x(\tau)$ for the free field and for the free detector observable that couples to the field 
\begin{eqnarray}
 C_\phi(x(\tau), x(\tau '))&=&\frac{1}{2}\<0|\{\phi_0(x(\tau)), \phi_0 (x(\tau'))\}|0\> \nonumber\\
 C_i(\tau, \tau ')&=&\frac{1}{2}\<E_i|\{M_0(\tau), M_0(\tau')\}|E_i \> \nonumber\\
 \chi_\phi(x(\tau), x(\tau '))&=&\frac{1}{2}\<0|[\phi_0(x(\tau)), \phi_0 (x(\tau'))]|0\>\nonumber\\
 \chi_i(\tau, \tau ')&=&\frac{1}{2}\<E_i|[M_0(\tau), M_0(\tau')]|E_i\>.
\end{eqnarray}
For a model with discrete energy levels we find that
\begin{eqnarray}
C_k(\tau, \tau ')&=&\sum_j |\<E_j|M_0(0)|E_k\>|^2 \cos((E_j -E_k)\Delta\tau)\nonumber\\
\chi_k(\tau, \tau ')&=&-\sum_j |\<E_j|M_0(0)|E_k\>|^2i\sin((E_j -E_k)\Delta\tau).
\end{eqnarray}
In addition, we may obtain the two-point functions for the scalar field via the relations $C_\phi(x,x')=\frac{1}{2} ({\cal G}^+(x,x') +{\cal G}^+(x',x))$ and $ \chi_\phi(x,x')= \frac{1}{2}({\cal G}^+(x,x')-{\cal G}^+(x',x))$.

\subsection{Vacuum Fluctuations and Radiation Reactions in D-dimensional Anti-de Sitter Spacetime}
With the formalism in place to analyze the excitations of the detector in detail, we make use the Wightman function derived in section \ref{wight} to calculate the vacuum fluctuations and radiation reactions that the detector experiences as it moves through the spacetime. The notion of a well defined temperature is of course restricted to time-scales over which the acceleration of the detector does not change appreciably. The vacuum fluctuations are obtained from
\begin{eqnarray}
\<0,E_i| \dot{H}(\tau)_{\mbox{\tiny vac }}|0,E_i \>&=& 2ic^2\int^\infty_{-\infty}d\tau 'C_{\phi}(x(\tau) , x(\tau '))\frac{d}{d\tau}\chi_i (\tau, \tau'). \nonumber\\
\end{eqnarray}
As before we have for the constant acceleration trajectories that $ C_{\phi}(\Delta \tau )$ is purely a function of the proper time difference between the two points.
The calculations are similar to those of the previous section and the vacuum fluctuations for the supercritical trajectories are found to be
\begin{eqnarray}
 \<0,E_i|\dot{H}(\tau)_{\mbox{\tiny vac}}|0,E_i \>&=&-2c^2\sum_{E_{ji}<0} |M_{ji}|^2 |E_{ji}| F (|E_{ji}|,a,k )\left [ 1 + \frac{(-1)^D + 1}{e^{2 \pi \frac{|E_{ji}|}{ \omega} } - (-1)^D} \right ] \nonumber \\
\nonumber \\
&& +2c^2\sum_{E_{ji}>0} |M_{ji}|^2 |E_{ji}|F ( |E_{ji}|,a,k) \left [ 1 + \frac{(-1)^D +1}{e^{2 \pi \frac{|E_{ji}|}{ \omega} } - (-1)^D} \right ]  \nonumber \\
\end{eqnarray}
with $E_{ji}= E_j -E_i$ and  $M_{ji} = \<E_j|M_0(0)|E_i \>$. We see that vacuum fluctuations as defined are unaffected in odd dimensional Anti-de Sitter and enhanced in even dimensional Anti-de Sitter. Naturally, this behaviour will also apply to Minkowski spacetime, which is obtained as $k \rightarrow 0$.

The radiation reactions are calculated similarly and we find that along the supercritical trajectories they are given by  
\begin{eqnarray}
 \<0,E_i|\dot{H}(\tau)_{\mbox{\tiny rad}}|0,E_i \>&=& -2c^2\sum_{E_{ji}<0} |M_{ji}|^2 |E_{ji}| F (|E_{ji}|,a,k )\left [ 1 + \frac{(-1)^D - 1}{e^{2 \pi \frac{|E_{ji}|}{ \omega} } - (-1)^D} \right ] \nonumber\\
&& -2c^2\sum_{E_{ji}>0} |M_{ji}|^2 |E_{ji}|F (|E_{ji}|,a,k ) \left [ 1 + \frac{(-1)^D -1}{e^{2 \pi \frac{|E_{ji}|}{ \omega} } - (-1)^D} \right ]. 
\end{eqnarray}

In addition to a response function `inversion of statistics' in odd dimensions we find that the vacuum fluctuations and radiation reactions, as defined above, are effectively swapped in odd dimensions. This is not surprising as it has been shown \cite{ooguri} that in Minkowski spacetime the unusual behaviour of two-point functions in odd dimensions is due to how the support of two-point functions in $D$ dimensions differs from the support in $D+1$ dimensions. For example for two spatial dimensions Huygen's principle does not hold for bosonic fields and `wave diffusion' occurs \cite{nash}.

The net effect of all of this is that the total rate of change of the detector's energy is given by
\begin{eqnarray}\label{total}
 \<0,E_i|\dot{H}(\tau)_{\mbox{\tiny total}}|0,E_i \>&=&-2c^2\sum_{E_{ji}<0} |M_{ji}|^2 |E_{ji}|F (|E_{ji}|,a,k ) \left [\frac{e^{2 \pi \frac{|E_{ji}|}{ \omega} }}{e^{2 \pi \frac{|E_{ji}|}{ \omega} } - (-1)^D} \right ] \nonumber \\
&& +2c^2\sum_{E_{ji}>0} |M_{ji}|^2 |E_{ji}|F (|E_{ji}|,a,k ) \left [ \frac{1}{e^{2 \pi \frac{|E_{ji}|}{ \omega} } - (-1)^D} \right ].
\end{eqnarray}
It is clear from this that an accelerated detector in its ground state will always experience excitations.
\subsection{Characteristic Proper Time and Detector Equilibrium}
It is useful to obtain a time-scale over which any heating from this coupling takes place. To achieve this we may consider the case of a `qubit detector' with two energies $\pm\frac{1}{2}\Delta E$ and $M_{ij}$ being a symmetric matrix.

Then for the system being in a general state, equation (\ref{total}) may be written as
\begin{eqnarray}
 \partial_t\<H\>&=& \frac{-\frac{1}{4}c^2 \Delta E F(\Delta E,a,k)}{e^{2\pi \frac{\Delta E}{\omega }}- (-1)^D}\left [ (e^{2 \pi \frac{\Delta E}{\omega }} -1) + ( e^{2 \pi \frac{\Delta E}{\omega }} +1) \frac{\<H(\tau )\>}{\frac{1}{2}\Delta E} \right ].   
\end{eqnarray}
The solution of which is
\begin{eqnarray}
 <H(\tau)>&=& E_e - (E_e - \<H(0)\>)e^{-\frac{\tau}{\tau_e}}
\end{eqnarray}
with
\begin{eqnarray}
 \tau_e&=& \frac{e^{2 \pi \frac{\Delta E}{\omega }} - (-1)^D}{\frac{1}{2}c^2 F(\Delta E, a,k) (e^{2 \pi \frac{\Delta E}{\omega }} +1)}  \non
E_e &=& -\frac{1}{2}\Delta E \tanh(\frac{\pi \Delta E}{\omega }).    
\end{eqnarray}
The characteristic time for the system to evolve into its equilibrium state with the scalar field is given by $\tau_e$, where $\tau_e =\frac{2}{c^2F}$ in odd dimensions and $\tau_e =\frac{4|E_e|}{c^2F\Delta E}$ in even dimensions. Futhermore, we see that the various factors such as the spacetime dimension and the curvature only affect the rate at which the detector approaches its equilibrium state - in all cases detailed balance is respected and the equilibrium state is thermal at the correct temperature $T=2\pi/\sqrt{a^2-k^2}$.

\section{Gibbons-Hawking temperature of de Sitter spacetime via an embedding in Anti-de Sitter spacetime}\label{GHtemp}

Having analyzed in some detail the behaviour of a detector in $D$-dimensional Anti-de Sitter spacetime, and deduced Unruh responses for accelerations above the curvature scale, we now present an application of these findings.

Previously \cite{gems2,gems3, hawking-unruh-equiv,levin} work has been done showing that one may derive thermal properties of particle production in a curved spacetime, such as 4D Schwarzschild spacetime, by a Global Embedding Minkowski Spacetime (GEMS) approach \cite{gems1}, in which a particle detector is constrained to a curved sub-manifold. In the case of 4D Schwarzschild the GEMS is 6-dimensional Minkowski, and in this embedding spacetime the detector accelerates and experiences a Minkowski Unruh response that coincides with the thermal response in the Schwarzschild spacetime.

In this section we show that it is also possible to relate gravitational particle creation in one spacetime to an Unruh response in a higher dimensional \emph{curved} spacetime. Specifically we embed a de Sitter spacetime in a higher dimensional Anti-de Sitter spacetime and, by considering a particle detector constrained to the sub-manifold, deduce the standard thermality of the expanding de Sitter spacetime.
 
Since we want to consider a comoving detector in the de Sitter spacetime as an accelerated detector in an embedding Anti-de Sitter spacetime, we must determine what is the correct AdS acceleration for the detector. One method is to consider a 4-D infinitely thin `brane' with constant tension $\sigma$ embedded in 5-D Anti-de Sitter spacetime (for related work on thermality in brane models see \cite{russo1, russo2}). The induced cosmological expansion on the 4D spacetime is determined by the Darmois-Israel junction conditions \cite{darmois, israel} across the brane. These state that the jump in the extrinsic curvature, $[K_{\mu \nu}]$, obeys the equation
\begin{eqnarray}
[K_{\mu \nu}] = - \kappa ^2 ( T_{\mu \nu} -\frac{1}{3} h_{\mu \nu} T) 
\end{eqnarray}
where $T_{\mu \nu}$ is the delta function energy tensor for the brane, $\kappa^2$ is the coupling of matter to gravity, and $h_{\mu \nu}$ is the induced metric on the lower dimensional spacetime. 

We let $u^\mu$ be the velocity vector of a comoving detector, and consider the tangential component of the extrinsic curvature, along its trajectory and define $K_{||}=u^\mu u^ \nu K_{\mu \nu}$. A direct calculation shows that $K_{||}=-n_\nu a^\nu$ where $a^\nu = u^\mu \nabla_\mu u^\nu$ is the acceleration of the comoving detector and $n^\mu$ is a unit vector field normal to the brane. Since we have a co-dimension one embedding, $n^\nu$ is uniquely determined up to a sign, and we so deduce that $K_{||}=-a$.

For a $T_{\mu \nu}$ describing a perfect fluid of density $\rho$ and pressure $p$, in the presence of a constant tension $\sigma$, we find that the junction condition implies that the acceleration of a comoving trajectory is given by $a=\frac{\kappa ^ 2}{6}( 2 \rho + 3 p - \sigma  )$. 

It is also possible to obtain the induced FRW equation for the embedded spacetime. To do so it is easiest to write the metric for the full Anti-de Sitter spacetime as
\begin{eqnarray}
ds^2&=& -f(r)dt^2 + \frac{dr^2}{f(r)} + r^2 d \Omega ^2_3
\end{eqnarray}
with $d \Omega ^2_3$ being the metric on a maximally symmetric 3-space that possesses a scale factor $r(\tau )$, and where $\tau $ is the proper time for a comoving observer in the lower dimensional spacetime. Using the spatial components of the junction conditions \cite{langlois}, we obtain
\begin{eqnarray}
H^2=\left (\frac{\dot{r}}{r}\right)^2&=& \frac{\kappa ^4}{36}\rho^2 + \frac{\kappa ^4 \sigma}{18}\rho+\frac{\kappa ^4 \sigma^2 }{36} -k^2.
\end{eqnarray}
For a de Sitter spacetime we thus require $\rho=p=0$, which gives an induced expansion rate given by $H^2 = \frac{\Lambda_4}{3}$ provided that we have the cosmological constant 
\begin{eqnarray}\label{lamb}
\Lambda _4=3\left (\frac{\kappa ^4 \sigma^2 }{36} -k^2 \right ) \ge 0.
\end{eqnarray}

We can then conclude that a particle detector moving with constant acceleration $a=-\kappa^2 \sigma /6$ in Anti-de Sitter will be constrained to a lower dimensional de Sitter spacetime with cosmological constant $\Lambda_4$ given by (\ref{lamb}). However, we have already shown that such a detector will have a thermal response with a temperature $T =\sqrt{a^2 -k^2} /2\pi$. From equation (\ref{lamb}) we deduce that $T=\sqrt{\Lambda_4/3}/2 \pi$, and we arrive at the Gibbons-Hawking temperature \cite{gibbons_hawking} measured by a comoving observer in the expanding de Sitter spacetime via the Unruh response in an embedding curved spacetime.

\section{Conclusion}\label{conclude}
In this paper we have considered an accelerated particle detector moving in D-dimensional Anti-de Sitter spacetime. In particular, our focus was to which aspects of the spacetime the detector is sensitive.

We derived a particularly useful form for a massless scalar field, and checked that it behaved correctly under certain limits. We then analyzed the detector response along the three different types of constant acceleration trajectories that occur in Anti-de Sitter. The accelerated particle detector does not suffer any response until its acceleration exceeds the curvature scale of the spacetime, at which point a modified thermal response occurs. The KMS condition holds for the detector at a temperature $T=\sqrt{a^2-k^2}/2\pi$, however in odd dimensions it exhibits a fermionic signature.

The response rate differs from a purely thermal one in all cases except for $k\rightarrow 0$ and $D=4$, the 4-dimensional Minkowski limit. Increasing the spacetime dimension produces polynomial modifications in $E/\sqrt{a^2-k^2}$, in addition to curvature dependent oscillations that weakly perturb the detector's response. 

We examined the response of the detector in some more detail by analysing its evolution in terms of vacuum fluctuations and radiation reactions. We demonstrated that the ground state of the detector was stable and evolved to the correct asymptotic thermal equilibrium state. In addition the vacuum fluctuations and radiation reactions are thermally modified in even and odd dimensions respectively.

Finally we demonstrated a generalization of the GEMS technique to a curved spacetime embedding. We considered de Sitter spacetime embedded in a larger dimensional Anti-de Sitter spacetime and showed that it was possible to derive the correct embedding acceleration via application of the junction conditions across a brane of constant tension $\sigma$. We showed that the cosmological constant on the brane was given by $\kappa^4 \sigma^2/12 -3k^2$, while a comoving trajectory corresponded to an acceleration $a=-\kappa^2 \sigma/6$ in the larger spacetime. As a result we arrived at the Gibbons-Hawking temperature $\sqrt{\Lambda_4/3}/2\pi$ via an Unruh response in the embedding Anti-de Sitter spacetime.

\end{document}